\documentclass[final]{WileyASNA-v1}

\articletype{Proceedings IWARA 2020}%

\received{}
\revised{}
\accepted{\hspace{325pt} IFT-UAM/CSIC-20-168}

\newcommand{\beq}{\begin{equation}}
\newcommand{\eeq}{\end{equation}}
\newcommand{\bp}{\boldsymbol{p}}

\begin{document}

\title{Reheating and Dark Matter Production}

\author[1]{Marcos A.~G.~Garcia}

\authormark{Marcos A.~G.~Garcia}

\address[1]{\orgdiv{Instituto de F\'{i}sica Te\'{o}rica (IFT) UAM-CSIC}, \orgaddress{\state{Madrid}, \country{Spain}}}

\corres{C/ Nicol\'as Cabrera 13-15, Campus de Cantoblanco UAM, 28049 Madrid, Spain.  \email{marcosa.garcia@uam.es}}

\abstract{A concise summary of out-of-equilibrium dark matter production during post-inflationary reheating is presented. We show that the dark matter relic abundance is in general sensitive to the thermalization rate of the inflaton decay products, and the evolution of the temperature of the subsequently thermalized radiation. We discuss how smoking-gun signals, such as monochromatic neutrinos or gamma-ray lines, or Lyman-$\alpha$ data, can help constrain out-of-equilibrium DM models. }

\keywords{Dark matter, inflation, reheating, indirect detection.}

\maketitle

\section{Introduction}

Since its inception~\citep{Gunn:1978gr}, the Weakly Interacting Matter Particle (WIMP) paradigm has been one of the most favored proposals to model the nature of dark matter (DM). This is in part to the so-called ``WIMP miracle'': electroweak-strength couplings between DM and Standard Model (SM) particles are of the right magnitude to have populated the Universe with the observed amount of DM via scatterings. DM would be produced in a state of thermal equilibrium with the background radiation, and therefore the correct abundance would be guaranteed regardless of initial conditions. The measured relic abundance would be reached after DM freezes-out of thermal equilibrium. Nevertheless, the WIMP paradigm appears to be in tension with direct detection experiments, which are continuously approaching the neutrino floor~\citep{Billard:2013qya}.
In part for these reasons, Feebly Interacting Matter Particle (FIMP) DM candidates have flourished~\citep{Hall:2009bx}. FIMPs are assumed to be part of a highly secluded ``dark sector'' that communicates with the SM through very suppressed
interactions. DM is therefore not produced while in thermal equilibrium. After decoupling, the DM momentum distribution would simply redshift due to the expansion of the Universe. Hence, at production, DM would take a ``snapshot'' of the extreme conditions of the early Universe, before recombination, and preserve it until much later times. In many FIMP constructions, DM could have been produced at the earliest possible times, during inflationary reheating. Here we provide a birds-eye overview of DM production during reheating, and discuss indirect detection signals and Lyman-$\alpha$ forest constraints for a specific DM model, namely a decaying non-supersymmetric spin-3/2 DM particle~\citep{Garcia:2020hyo}.

\section{Reheating}

The end of inflation is marked by the end of the accelerated expansion of the Universe, $\ddot{a}=0$, where $a$ denotes the scale factor. For inflation driven by the classical rollover of a scalar field $\Phi$ in its potential $V(\Phi)$, the end of inflation occurs when $\dot{\Phi}^2=V(\Phi)$. Once this occurs, the inflaton proceeds to perform damped oscillations about its minimum at $\Phi=0$. Simultaneously, the interaction between $\Phi$ and ordinary matter and radiation drives the conversion of the inflaton energy density and hence the reheating of the Universe. This reheating process is necessarily model-dependent. The damping rate, quantified by the Hubble parameter $H=\dot{a}/a$, and therefore the time-dependence of the inflaton oscillation, are a function of the shape of the potential. Moreover, the rate at which the thermal bath is populated depends on the specific form of the coupling of the inflaton to the visible and dark sectors. 

In the perturbative regime, the dynamics of the inflaton-plasma system are determined by the coupled Friedmann-Boltzmann system of the form
\begin{align} \displaybreak[0] \label{eq:FB1}
\dot{\rho}_{\Phi} + 3H (\rho_{\phi}+P_{\phi}) \;&=\; -\Gamma_{\Phi} \rho_{\Phi}\,,\\ \displaybreak[0] \label{eq:FB2}
\dot{\rho}_R + 4H\rho_R \;&=\; \Gamma_{\Phi}\rho_{\Phi}\,,\\ \displaybreak[0] \label{eq:FB3}
\rho_{\Phi} + \rho_R \;&=\; 3M_P^2 H^2\,,
\end{align}
where $\rho_{\Phi} \;=\; \frac{1}{2}\dot{\Phi}^2 + V(\Phi)$ and $P_{\Phi} \;=\; \frac{1}{2}\dot{\Phi}^2 - V(\Phi)$ denote the energy and pressure densities of the oscillating inflaton, respectively, $\rho_R$ is the energy density of its relativistic decay products, and $M_P=(8\pi G)^{-1/2}$ is the reduced Planck mass. For the sake of generality, consider a power-law form for the potential about the minimum,
\beq\label{eq:potk}
V(\Phi) \;=\; \lambda \frac{|\Phi|^k}{M_P^{k-4}}\,.
\eeq
This is, for example, the small field limit of T-attractor constructions~\citep{Kallosh:2013hoa}. For a quadratic minimum, the resulting oscillation of $\Phi$ is harmonic in time-scales shorter than the inverse damping rate. For $k>2$, the oscillations are anharmonic. We are interested in particle production that occurs in time-scales longer than the typical oscillation time.\footnote{We do not consider here particle production at short time-scales, which is a consequence of the violation of adiabaticity as the inflaton changes signs during its oscillation~\citep{Dolgov:1989us,Kofman:1994rk}.} It is straightforward to verify that for a potential of the form (\ref{eq:potk}), averaging over one oscillation results in the relation $\langle P_{\phi}\rangle = \frac{k-2}{k+2}\langle \rho_{\Phi}\rangle$~\citep{Turner:1983he}. Eq.~(\ref{eq:FB1}) can then be rewritten as
\beq
\dot{\rho}_{\Phi} + 3\left(\frac{2k}{k+2}\right)H\rho_{\phi} \;=\; -\Gamma_{\Phi}\rho_{\Phi}\,.
\eeq

The form of the inflaton potential also determines the time-dependence of the dissipation rate $\Gamma_{\Phi}$. This rate is not identical to the width of a single inflaton particle decaying in the vacuum, as reheating is driven by the collective decay of a classical, spatially homogeneous, oscillating condensate. Nevertheless, up to $\mathcal{O}(1)$ factors, which can be absorbed by redefining effective couplings, the dissipation rate can be recovered from the familiar computation of the particle decay rate, upon the substitution of the effective inflaton mass, defined as follows, 
\beq
m_{\Phi}^2(t) \;\equiv\; \partial_{\Phi}^2 V(\Phi)\big|_{\Phi=\Phi_0} \;=\; k(k-1)M_P^{\frac{2(4-k)}{k}}\lambda^{\frac{2}{k}}\rho_\Phi^{\frac{k-2}{k}}\,.
\eeq
Here $\Phi_0$ denotes the instantaneous value of the decaying envelope of the oscillating inflaton, given by the relation $\rho_{\Phi}=V(\Phi_0)$~\citep{Shtanov:1994ce}. As an example, consider that the dominant decay channel of the inflaton is to a pair of fermions with coupling $y\,\Phi\bar{f} f$. The time-dependent dissipation rate will then be given by
\beq\label{eq:gammafer}
\Gamma_{\Phi} \;=\; \frac{y^2}{8\pi}m_{\Phi}(t)\,.
\eeq
The effective width will have a different energy density dependence if the dominant decay channel proceeds is determined by a different coupling, such as one to spin-0 bosons~\citep{Ichikawa:2008ne,Kainulainen:2016vzv}.

The system of equations (\ref{eq:FB1})-(\ref{eq:FB3}) with rate (\ref{eq:gammafer}) can be solved in an approximate analytical form for $t< t_{\rm reh}$, where $t_{\rm reh}$ denotes the time of inflaton-radiation equality~\citep{Garcia:2020eof},
\begin{align}
\rho_{\Phi} \;&\simeq\; \rho_{\rm end}\left(\frac{a}{a_{\rm end}}\right)^{-\frac{6k}{k+2}}\,,\\
\rho_R \;&\simeq\; \frac{y^2}{8 \pi}\sqrt{3k(k-1)}\lambda^{\frac{1}{k}}
M_P^{\frac{4}{k}} 
\left(\frac{k+2}{14-2k} \right)
\nonumber
\\
&\qquad\times \rho_{\rm end}^{\frac{k-1}{k}}
\left( \frac{a_{\rm end}}{a} \right)^4 
\left[
\left(\frac{a}{a_{\rm end}} \right)^{\frac{14-2k}{k+2}}-1
\right]\,.
\end{align}
Here the subindex ``end'' denotes quantities evaluated at the end of inflation. For $a\gg a_{\rm end}$, $a\propto t^{\frac{k+2}{3k}}$. At later times, $t\gtrsim t_{\rm reh}$, it can be shown that the decay of the inflaton is exponential in time, $\rho_{\Phi}\propto e^{-\Gamma_{\Phi} t}$, for $k=2$, while for $k>2$ the decay is power-like in time, $\rho_{\Phi}\propto t^{\frac{2k}{2-k}}$~\citep{GKMO}.

\begin{figure}[t!]
\includegraphics[width=1.0\columnwidth]{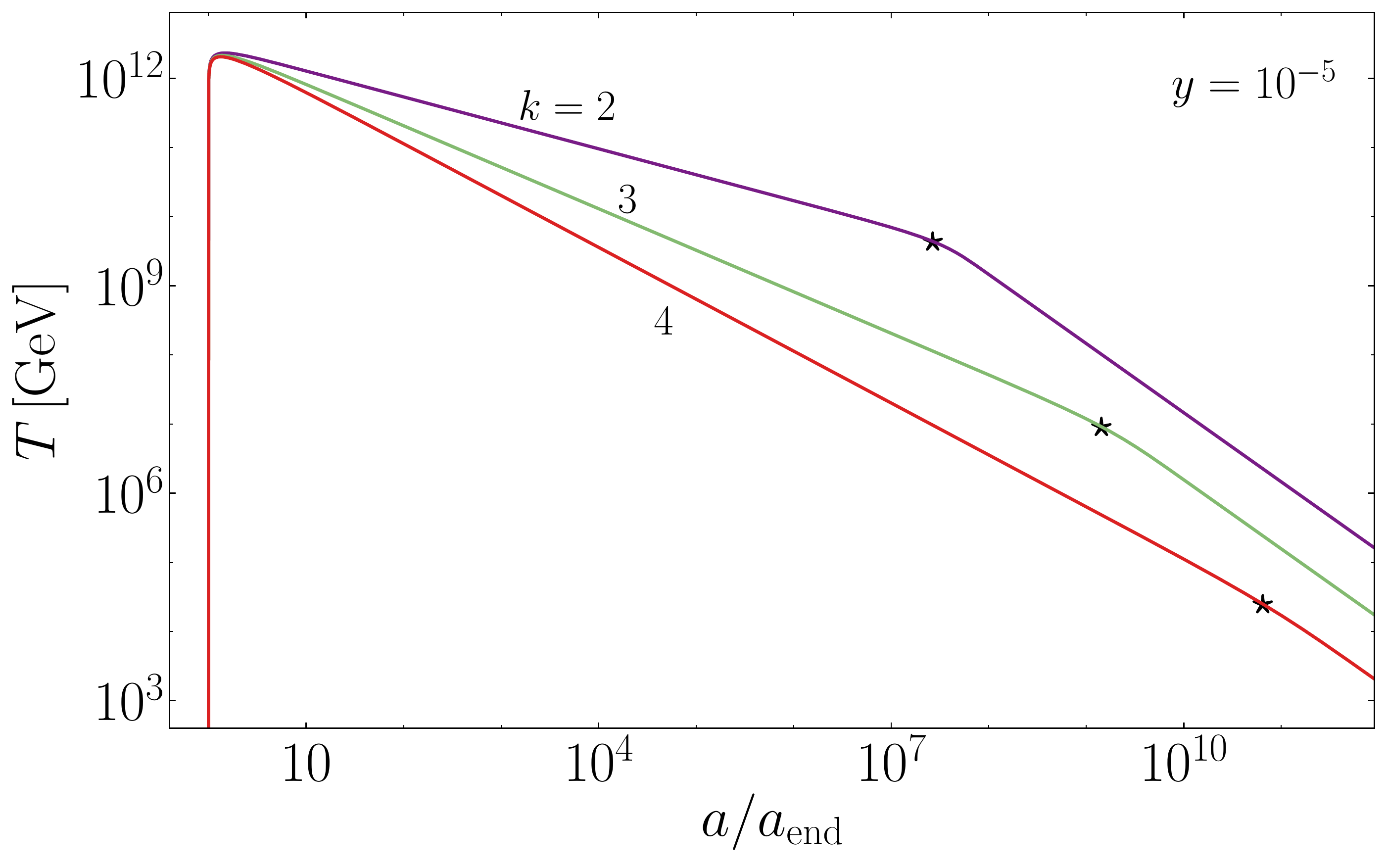}
\caption{Scale-factor dependence of the instantaneous temperature during reheating for selected values of $k$ with $y=10^{-5}$. The star signals inflaton-radiation equality, $\rho_{\Phi}=\rho_R$, corresponding to $T=T_{\rm reh}$.
}
\label{fig:TgPlotA}
\end{figure}

After they are created, the relativistic decay products of the inflaton interact with each other, eventually reaching the state of thermal equilibrium. When that occurs, their temperature is simply related to their energy density via the usual thermodynamic relation,
\beq
T \;=\; \left(\frac{30 \rho_R}{\pi^2 g_*} \right)^{1/4}\,,
\eeq
with $g_*$ the number of relativistic degrees of freedom. Assuming thermalization is instantaneous during reheating, this temperature is affected by the cooling due to expansion, and by the heating due to inflaton decay. Fig.~\ref{fig:TgPlotA} shows the evolution of the temperature with the scale factor for $k=2,3,4$ and $y=10^{-5}$. When inflation ends, the temperature sharply rises to a maximum value, which we denote $T_{\rm max}$. This maximum temperature is a weak function of the potential shape. However, at later times, a drastically different thermal history is obtained depending on the value of $k$, with
\beq
T \;\propto\; a^{-\frac{3k-3}{2k+4}}\,.
\label{scale}
\eeq
before reheating ends when the Universe becomes dominated by radiation. Note in particular that the temperature at the end of reheating, $T_{\rm reh}$, is much lower for larger $k$~\citep{Garcia:2020eof}.

In a realistic scenario, thermalization is not instantaneous. The inflaton decay products are initially produced at very high energies, with momenta $p=\mathcal{O}(m_{\Phi})$. It is not until elastic and inelastic scatterings become efficient relative to the expansion rate that kinetic and chemical equilibrium can be established. For a quadratic minimum, the thermalization time-scale is known~\citep{Harigaya:2013vwa,Mukaida:2015ria}. For processes mediated by a gauge interaction with coupling $\alpha$, small angle $2\rightarrow 3$ inelastic splittings rapidly (over)populate the primordial plasma with a bath of soft particles of momentum $p\ll m_{\Phi}$. These soft particles eventually produce a thermalized background that is capable of cooling down the hard, energetic inflaton decay products. The hard non-thermal sector is completely depleted, and thus thermalization achieved, at
\beq\label{eq:tth}
t_{\rm th} \;\simeq\; \Gamma_{\Phi}^{-1}\alpha^{-16/5} \left(\frac{\Gamma_{\Phi}m_{\Phi}^2}{M_P^3}\right)^{2/5}\,,
\eeq
which for perturbative reheating is generically $t_{\rm th}\ll t_{\rm reh}$. At this moment of time the true maximum temperature of the Universe is reached, 
\beq
T_{\rm th} \;\simeq\; \alpha^{4/5}m_{\Phi}\left(\frac{24}{\pi^2g_{\rm reh}}\right)^{1/4}\left(\frac{\Gamma_{\Phi}M_P^2}{m_{\Phi}^3}\right)^{2/5}\,.
\eeq

\section{Dark Matter Production}

We now proceed to apply the picture of perturbative reheating presented above to the out-of-equilibrium production of dark matter. In freeze-out models, DM is assumed to interact sufficiently strongly at early times, being kept in thermal equilibrium with the SM. On the contrary, for out-of-equilibrium (freeze-in) production, DM couplings are assumed to be so suppressed that DM never reaches thermal equilibrium, with itself or the plasma~\citep{McDonald:2001vt,Hall:2009bx}. 

Specializing to DM production from $2\rightarrow 2$ scatterings, the above mentioned suppression can be originated from the presence of heavy mediators, couplings suppressed by higher order operators, or a combination of both. Generically, the integrated effective cross section of the production process can be written in the form
\beq
\sigma(s) \;\propto\; \frac{s^{\frac{n}{2}}}{\Lambda^{n+2}}\,,
\eeq
where $s=\sqrt{E}$ denotes the center of mass energy of the scattering, and $\Lambda$ can be thought to be parametrically related to the mass of a heavy mediator in the UV-theory. For non-negative $n$, the suppression by $\Lambda$ guarantees that the DM abundance is determined by forward processes (plasma $\rightarrow$ DM) rather than by annihilations. Pauli-blocking/Bose-enhancement can be safely disregarded, and DM never reaches thermal equilibrium. Moreover, assuming no post-reheating entropy production, DM production is dominated at temperatures $T\geq T_{\rm reh}$~\citep{Elahi:2014fsa,Garcia:2017tuj}. 

\begin{figure*}[t]
\centering
    \subfloat[]{\includegraphics[width=0.48\linewidth]{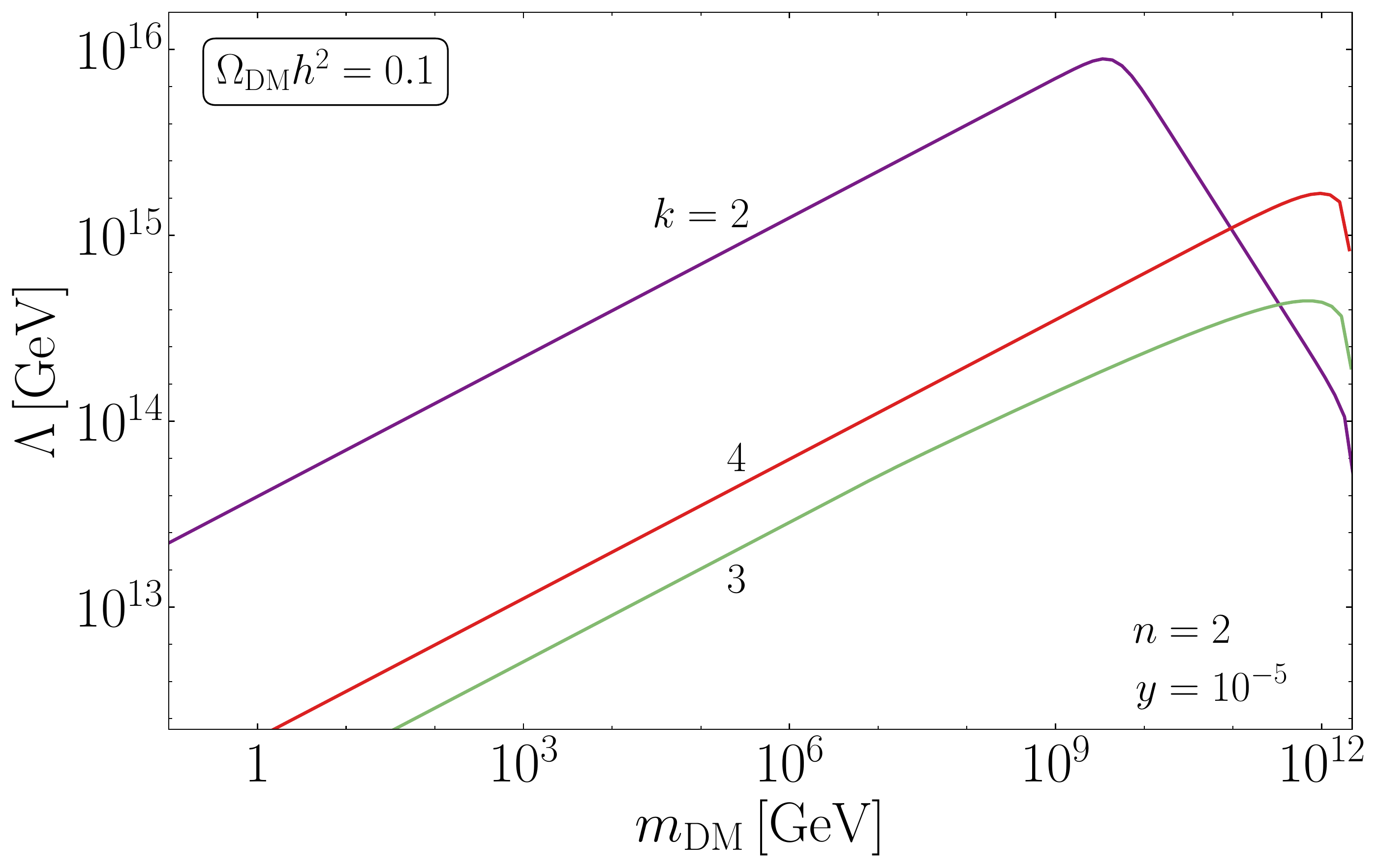}\label{sub:a}}
\hfill
    \subfloat[]{\includegraphics[width=0.48\linewidth]{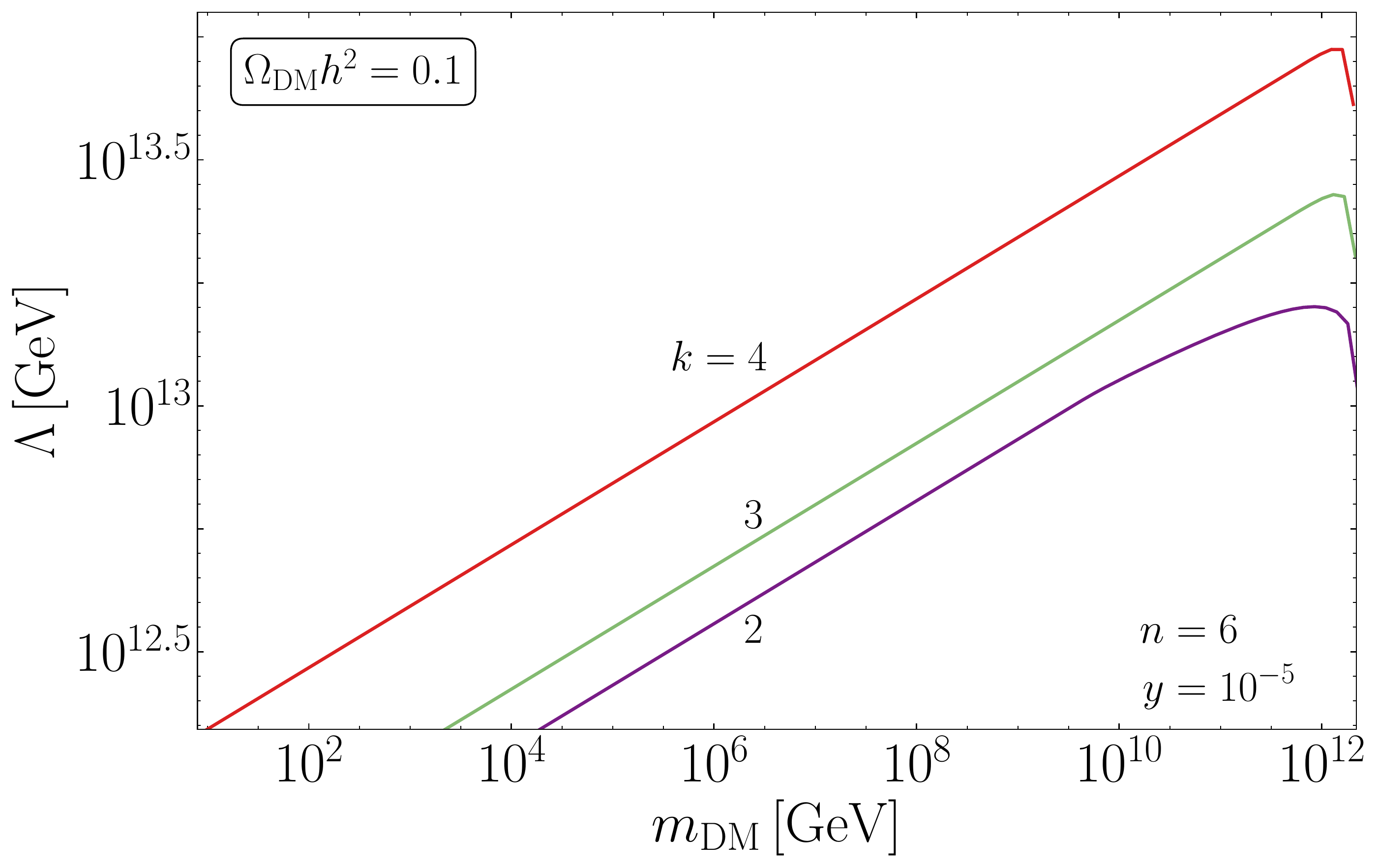}\label{sub:b}}
\caption{Contours $\Omega_{\rm DM}h^2=0.1$ showing the required value of $\Lambda$ as a function of the DM mass. We assume an inflaton decay coupling $y=10^{-5}$ and a production rate with $n=2$ (left) and $n=6$ (right). }
\label{fig:GenFern}
\end{figure*}
Given a (freeze-in) DM model, the determination of the corresponding relic abundance follows a simple recipe: one needs to integrate the Boltzmann transport equation, disregarding DM annihilations and statistical Bose enhancement / Pauli blocking effects, which are suppressed. For the DM number density $n_{\chi}$, the Boltzmann equation takes the form
\beq\label{eq:boltzn}
\dot{n}_{\chi} + 3H n_{\chi} \;=\; R(t)\,,
\eeq
where the production rate of DM (per unit volume per unit time) is given by
\beq
R(t) \;=\; 2\int \frac{d^3 \bp_1}{(2\pi)^3 2p_1^0}\frac{d^3 \bp_2}{(2\pi)^3 2p_2^0}s\,\sigma(s) f_1(\bp_1,t) f_2(\bp_2,t)\,.
\eeq
Here $\bp_{1,2}$ denote the momenta of the incoming scatterers, produced from the decay of the inflaton, and $f_{1,2}$ their corresponding phase space distributions. The evolution of these distributions is highly non-trivial. As discussed previously, the inflaton decay products are initially in a non-thermal state. For $t\ll t_{\rm reh}$, the distribution of these products is given by~\citep{McDonald:1999hd,Garcia:2018wtq}
\beq
f_{1,2}(\bp,t) \;\simeq \; \frac{24\pi^2 n_{1,2}(t) }{g_{\chi} m_{\Phi}^3} \left(\frac{m_{\Phi}}{2p}\right)^{3/2} \, \theta(m_{\Phi}/2-p)\,,
\eeq
corresponding to a spectrum peaked at $p=m_{\Phi}/2$. Only after complete thermalization has been reached these distributions take their usual Fermi-Dirac or Bose-Einstein form, $f_{1,2}(\bp,t)=(1\pm e^{p/T(t)})^{-1}$.

Assuming that neither DM nor the parent scatterers that annihilate to produce DM are directly produced by the decay of the inflaton, DM will only be produced after thermalization. Integration of Eq.~(\ref{eq:boltzn}) is then relatively straightforward, as for a general inflaton potential the time-dependence of the temperature during reheating is simply given by Eq.~(\ref{scale}).  If the reheating temperature is always larger than the DM mass, and with $R(T)=T^{n+6}/\Lambda^{n+2}$, the resulting expression for the closure fraction $\Omega_{\chi}=\rho_{\chi}/\rho_{\rm tot}$ has the following form~\citep{Garcia:2020eof}
\begin{align}\notag \displaybreak[0]
&\Omega_{\rm DM}h^2 \;\simeq\; 1.4 \times 10^{2-6n} \left(\dfrac{427/4}{g_{*,{\rm reh}}}\right)^{3/2} \left(\dfrac{m_{\rm DM}}{1 \,{\rm GeV}}\right) \\ \notag \displaybreak[0]
&\ \times \left(\dfrac{10^{16}\,{\rm GeV}}{\Lambda}\right)^{n+2} 
\left(\dfrac{T_{\rm reh}}{10^{10}\,{\rm GeV}}\right)^{n+1} \\ \label{ee:omegaTgen}
&\ \times \begin{cases}
\dfrac{2k+4}{n-nk+10-2k}  \,,\quad & n<\dfrac{10-2k}{k-1}\,,\\[10pt]
\dfrac{2k+4}{k-1}  \ln\left( \dfrac{T_{\rm max}}{T_{\rm reh}}\right) \,,\quad & n=\dfrac{10-2k}{k-1} \,,\\[10pt]
\dfrac{2k+4}{nk-n+2k-10} \left(\dfrac{T_{\rm max}}{T_{\rm reh}}\right)^{\frac{2k-10}{k-1}+n}  \,,\ & n>\dfrac{10-2k}{k-1}\,.
\end{cases}
\end{align}
Note the dependence of this result on the shape of the inflaton potential, given by $k$, the coupling of $\Phi$ to matter, which determines $T_{\rm max}$ and $T_{\rm reh}$, and the production cross section, through $\Lambda$ and $n$. For a quadratic inflaton potential, with $k=2$, the bulk of the dark matter relic abundance will be produced at the {\it lowest }temperature during reheating, $T_{\rm reh}$, if $n<6$. This occurs since the dominant process that determines the evolution of the DM number density is the production of entropy from the decay of $\Phi$. This dilution mechanism suppresses the population of the Universe with a significant fraction of DM until the end of reheating. On the other hand, if $n\geq 6$, the closure fraction becomes dependent on the {\it highest} temperature during reheating, $T_{\rm max}$. In these cases, the cross section has such a steep dependence on the energy of the scatterers that the enhanced production at the end of inflation dominates over dilution due to the dissipation of $\Phi$. For a quartic potential, dominant production at $T_{\rm max}$ occurs for $n\geq 2/3$, which in practice implies that any production process with a cross section that increases with the temperature will populate the Universe with DM immediately after thermalization. The dependence of the relic abundance on $\Lambda$, $k$ and the DM mass is presented in Fig.~\ref{fig:GenFern}. We note the relevance of the potential shape, and the dependence on early dynamics. The change in the slope of the curves occurs when $m_{\rm DM}= T_{\rm reh}$. For larger DM masses, thermal production is effective only for $T_{\rm max}> T > m_{\rm DM}$ in this case, and the expression (\ref{ee:omegaTgen}) is modified (see~\citep{Garcia:2020eof}).

\begin{figure}[t!]
\includegraphics[width=1.0\columnwidth]{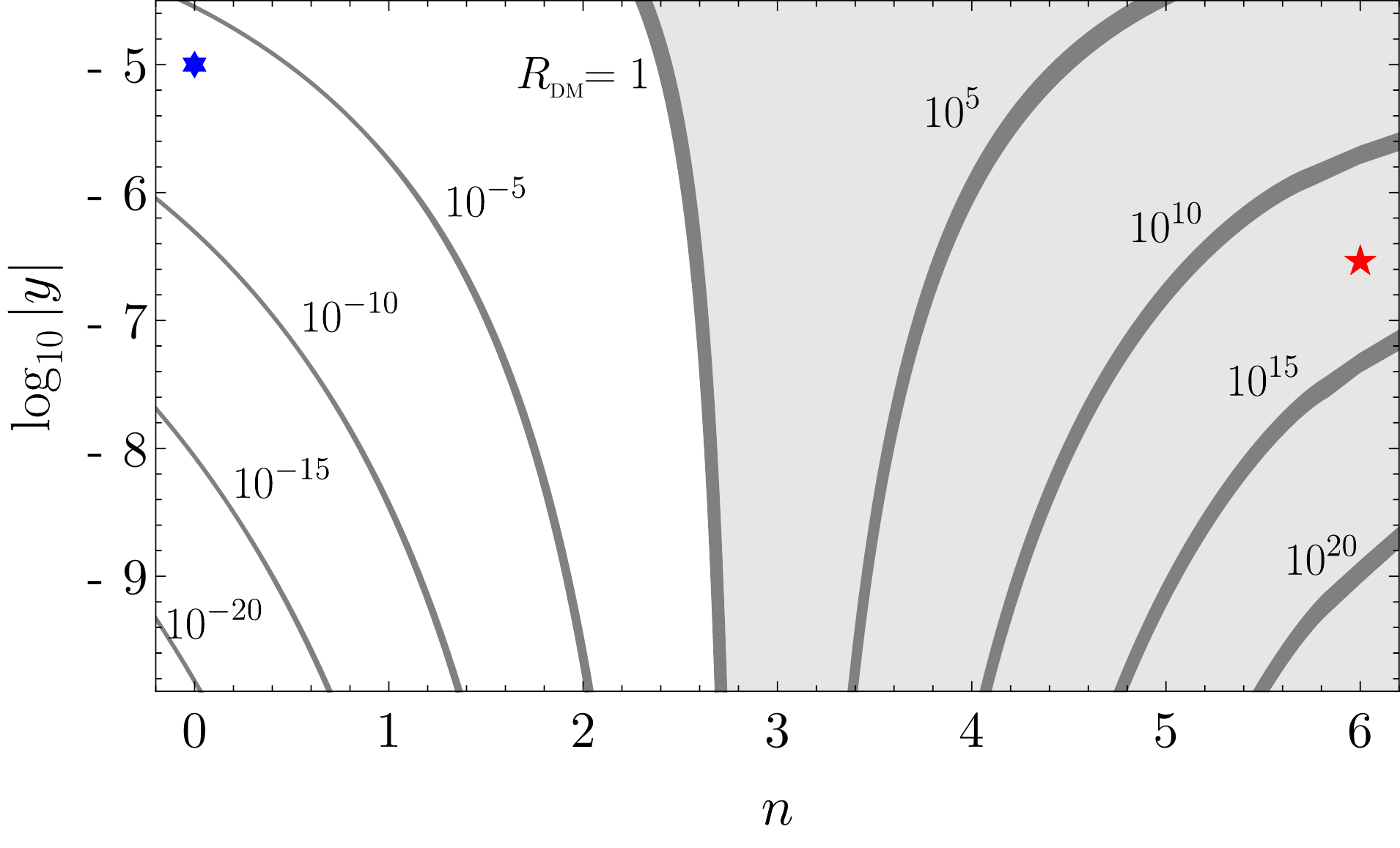}
\caption{Contours of the nonthermal to thermal yield ratio at the end of reheating, $R_{\rm DM}=Y^{\rm (NT)}_{\rm DM}/Y^{\rm (T)}_{\rm DM}$ , in the $(n,y)$ plane. Thermalization proceeds via the strong interaction. In the shaded region particle production is nonthermally dominated.
}
\label{fig:nonthermal}
\end{figure}

If the DM parent scatters are produced directly from the decay of the inflaton, DM may have been produced in the pre-thermal phase of reheating. For a quadratic inflaton potential, the closure fraction for these pre-thermal relics has been determined. Writing $R(t)=n_R^2(t) m_{\Phi}^n/\Lambda^{n+2}$, one obtains
\beq
\Omega_{\rm DM}h^2 \;\simeq\; 10^{27} \frac{(\Gamma_{\Phi}M_P)^{3/2}(\Gamma_{\Phi}t_{\rm th})m_{\Phi}^{n-2} m_{\rm DM}}{g_{*,{\rm reh}}^{1/4}\Lambda^{n+2}}\,,
\eeq
Comparison with (\ref{ee:omegaTgen}) reveals that the non-thermal abundance dominates over the thermally produced one if $n>2$~\citep{Garcia:2018wtq}. This result can be seen in Fig.~\ref{fig:nonthermal}, which shows the ratio of the nonthermal and thermal yields, $Y_{\rm DM}\equiv n_{\rm DM}/n_R$, as a function of the Yukawa coupling $y$ and $n$, assuming that the inflaton predominantly decays to the DM parent scatterers. The blue six-pointed star corresponds to gravitino production in low-scale supersymmetry breaking~\citep{Bolz:2000fu,Pradler:2006qh,Rychkov:2007uq,Eberl:2020fml}, while the red five-pointed star corresponds to the observed DM abundance for a DM gravitino in high scale supersymmetry~\citep{Benakli:2017whb,Dudas:2017rpa}. In this later case, for which $n=6$, the instantaneous thermalization approximation underestimates the DM closure fraction by a factor of $ \sim10^{14}$. This example demonstrates the sensitivity of the relic abundance to the thermalization time-scale and inflationary parameters such as $m_{\Phi}$, making DM an interesting probe for reheating dynamics.

\section{Constraining Out-of-equilibrium dark matter}

For DM to play the role of a probe of the dynamics that occurred in the early Universe, we must be able to detect it first. With the current experimental sensitivity, direct detection for these feeble interacting dark matter models is unlikely, if not impossible. Nevertheless, indirect signals can provide almost equally powerful insights. 

An out-of-equilibrium scenario which conveniently illustrates the roles that indirect detection, thermal and non-thermal effects, and constraints beyond the value of $\Omega_{\rm DM}$ have,  is given by the decaying, non-supersymmetric spin-3/2 DM model proposed in~\citep{Garcia:2020hyo}. The DM particle, dubbed the {\it raritron} ($\Psi_{\mu}$) couples to radiation, right-handed (sterile) neutrinos and matter as follows,
\begin{eqnarray}
{\cal L}_{3/2} & = & i\frac{\alpha_1}{2 M_P} \bar \nu_R \gamma^\mu [\gamma^\rho,\gamma^\sigma] \Psi_\mu F_{\rho \sigma}
+ {\rm h.c.} \nonumber \\
 & & + i\frac{\alpha_2}{{2}M_P} i \sigma_2 (D^\mu H)^* \bar L  \Psi_\mu  \, .
\label{Eq:l32}
\end{eqnarray}
Assuming that the coupling $\alpha_2$ is negligibly small, the relic abundance and the lifetime of $\Psi_{\mu}$ will be determined solely by $\alpha_1$. Depending on the raritron mass, the dominant decay channel may contain either two or three final states. The 2-body decay channel $\Psi_\mu \rightarrow \nu A_\mu$ is always available. For $m_{3/2} > m_Z$, the $\nu Z_\mu$ final state is open and the 2-body final state dominates for $m_{3/2} \lesssim 6$ TeV. When $m_{3/2} \gtrsim m_H (m_H + m_Z)$, the channel $\Psi_\mu \rightarrow \nu H A_{\mu} (Z_\mu)$ opens up. For $m_{3/2} \gtrsim 6$ TeV, the 3-body final state dominates the decay width. Depending on the raritron mass, gamma ray and neutrino monochromatic signals can  be produced, which are suitable for their detection at observatories such as FERMI or IceCube.

The raritron can be produced directly from the relativistic plasma during reheating, through the  out-of-equilibrium annihilation processes, $H + \nu \rightarrow B + \Psi_{\mu}$, $H + B \rightarrow \nu + \Psi_{\mu}$, and $B + \nu \rightarrow H + \Psi_{\mu}$. For all of them, $\sigma(s) \propto s^2$, that is, $n=4$. Hence, pre-thermal DM production may be the dominant channel. For definiteness, under the assumption that the inflaton decays primarily into Higgs bosons and subdominantly to neutrinos, DM is produced before thermalization if the inflaton to neutrino branching ratio is ${\rm Br}_{\nu} \gtrsim 7\times 10^{-4} (T_{\rm reh}/10^{10}\,{\rm GeV})^{6/5}$.

Raritron DM can in principle be very light, but not arbitrarily light. Light DM will in general be non-cold, with a non-vanishing free-streaming velocity that may interfere with the gravitational formation of structure. Beyond their free-streaming scale, structure would form as in the case of perfectly cold DM, whereas for smaller scales, the non-vanishing velocity dispersion would wash out structure. This effect appears in the matter power spectrum as a step-like suppression, which for thermal relics (Warm Dark Matter, WDM) occurs around the Mpc range for $m_{\rm DM}\sim {\rm keV}$. These scales correspond to the region probed by the Lyman-$\alpha$ forest observation of distant quasars~\citep{Boera:2018vzq}. The Lyman-$\alpha$ constraint on WDM is precisely $m_{\rm WDM}\gtrsim {\rm keV}$~\citep{Viel:2005qj,Palanque-Delabrouille:2019iyz,Garzilli:2019qki}.

\begin{figure}[t!]
\includegraphics[width=1.0\columnwidth]{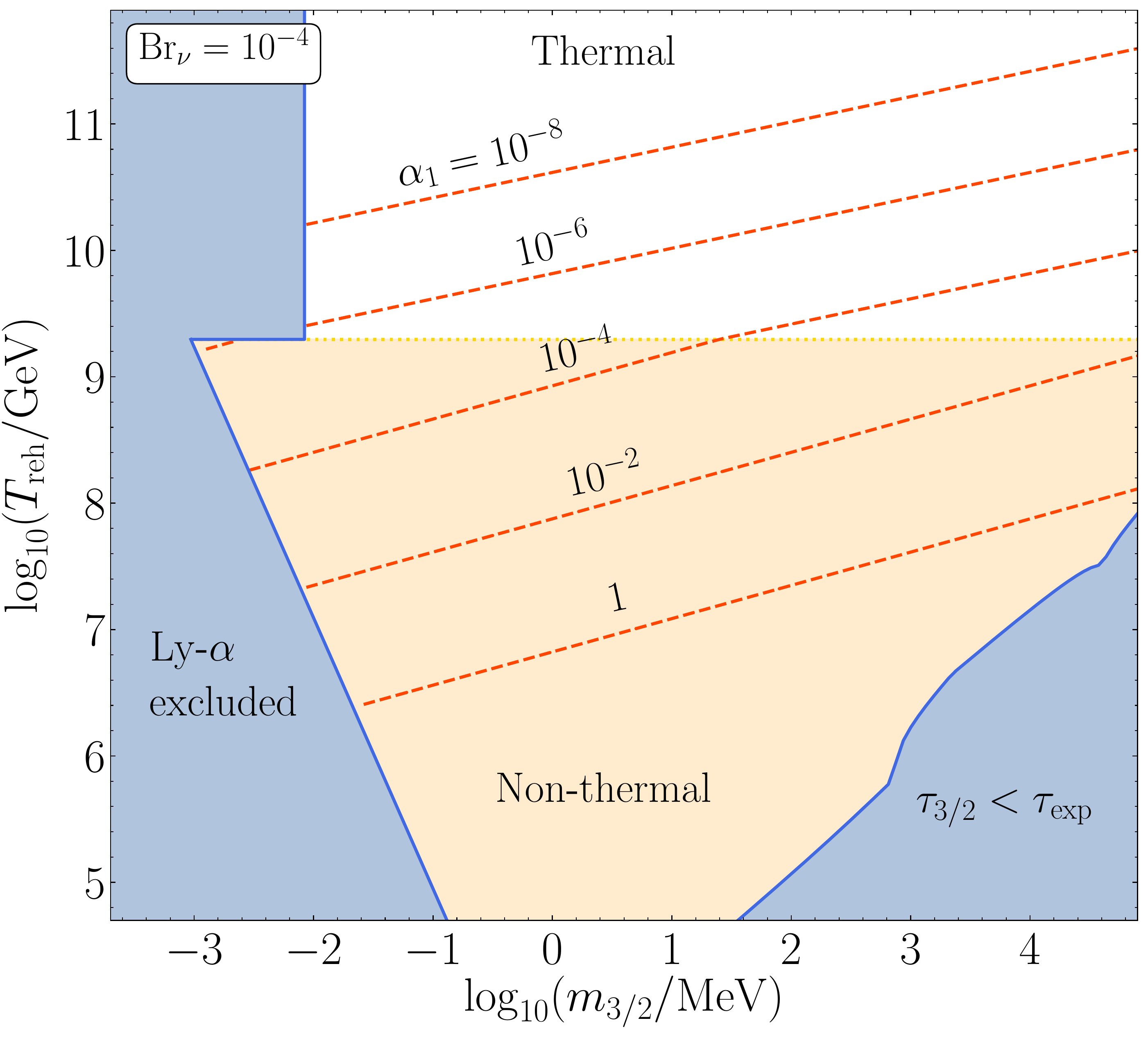}
\caption{$\Omega_{\rm DM}h^2=0.1$ curves in the ($m_{\rm DM}$, $T_{\rm reh}$) plane for the raritron model~(\ref{Eq:l32}), for different values of $\alpha_1$, and ${\rm Br}_{\nu}=10^{-4}$. The astrophysical constraints on the lifetime from $\gamma$-ray observations, and the Ly-$\alpha$ constraint on the mass are shown in blue. In the orange region freeze-in occurs non-thermally. 
}
\label{fig:raritron}
\end{figure}

A light raritron can be a WDM mimicker: it could be sufficiently relativistic during structure formation to leave a detectable imprint in the matter power spectrum at scales below their free-streaming horizon, yet heavy enough to be indistinguishable from cold dark matter at late times. For out-of-equilibrium DM, a general bound-mapping formalism has been devised in~\citep{BGP} (see also~\cite{Kamada:2019kpe}). If the phase space distribution of the DM is known, the Ly-$\alpha$ bound may be rescaled as
\beq
m_{\rm DM} \;\gtrsim\; 7.5~\text{keV}  \,  \left( \dfrac{m_{\rm WDM}}{3~\text{keV}} \right)^{4/3} \left( \dfrac{\langle p \rangle_{0}}{T_{0}} \right) \dfrac{\sqrt{\langle q^2 \rangle}}{\langle q \rangle}\,,
\eeq
where $\langle p \rangle_{0}$ and $T_0$ denote the mean DM momentum and the photon temperature at the present time, respectively, and $q$ is the ratio of the comoving DM momentum to a convenient energy scale that characterizes the DM density at the present time, denoted by $T_{{\rm NCDM},0}$ in the literature~\citep{Lesgourgues:2011rh}.  The resulting bounds depend on the reheating dynamics and the production mechanism, and can vary by many orders of magnitude. For the raritron, $m_{\rm DM} \gtrsim 8\,{\rm keV}$ in the case of thermal production, assuming $m_{\rm WDM}=3\,{\rm keV}$. For non-thermal production, $m_{\rm DM} \gtrsim 0.4\,{\rm keV}\left(10^{10}\,{\rm GeV}/T_{\rm reh}\right)^{7/15}$ with $\alpha=0.03$ (cf. Eq.~(\ref{eq:tth})). Fig.~\ref{fig:raritron} shows the cons\-traints on the model parameter space coming from Ly-$\alpha$ data, from $\gamma$-ray observations, and from the observed DM relic abundance. For low reheating temperatures, $T_{\rm reh}\lesssim 4\times 10^{9}\,{\rm GeV}$, and  ${\rm Br}_{\nu}=10^{-4}$, prethermal effects determine the value of $\Omega_{\rm DM}$.

\section{Outlook}

We have here motivated and discussed our recent efforts towards the characterization of out-of-equilibrium dark matter production during reheating. We have found that the evolution of energy densities, and therefore of the temperature of the primordial plasma, depend on the shape of the inflaton potential near its minimum. For a dark matter production rate with a strong temperature dependence, the relic abundance is therefore also dependent on this potential, and may be grossly underestimated if one fails to account for these effects. Moreover, for a quadratic inflaton potential, the thermalization time-scale of the relativistic decay products of the inflaton can be estimated. Again, if the production rate of dark matter depends steeply on the energy of the scattering parent particles, the overwhelming fraction of dark matter could be a relic of the pre-thermal Universe, providing a potential window to the inflation-reheating transition. In addition, we characterized a model of decaying spin-3/2 dark matter, for which astrophysical $\gamma$-ray and neutrino monochromatic signals are a generic prediction. For this example, we also illustrated the bounds that the measurement of the matter power spectrum from the Lyman-$\alpha$ forest can provide. Despite the breadth of our discussion, many more overlooked effects, such as non-perturbative particle production and thermal corrections, can make a difference in the determination of the primordial dark matter abundance. Hence,  much more work needs to be done in the study of the connection between reheating and dark matter. 

{\bf Acknowledgements.} M.A.G.G.~is supported by the Spanish Agencia Estatal de Investigaci\'on through the Grants No.~FPA2015-65929-P (MINECO/FEDER, UE) and No.~PGC2018095161-B-I00, IFT Centro de Excelencia Severo
Ochoa Grant No.~SEV-2016-0597, and Red Consolider MultiDark
Grant No.~FPA2017-90566-REDC.

\bibliography{biblio}%


\end{document}